# Growth of flat SrRuO$_3$ (111) thin films suitable as bottom electrodes in heterostructures

D. Rubi, A. Vlooswijk and B. Noheda

Zernike Institute for Advances Materials, University of Groningen, Nijenborgh 4, 9747AG, The Netherlands

Thin film growth of ferroelectric or multiferroic materials on SrTiO$_3$(111) with a buffer electrode has been hampered by the difficulty of growing flat electrodes on this polar orientation. We report on the growth and characterization of SrRuO$_3$ thin films deposited by pulsed laser deposition on SrTiO$_3$(111). We show that our SrRuO$_3$(111) films are epitaxial and display magnetic bulk-like properties. Films presenting a thickness between 20 and 30nm are found to be very flat (with an RMS of about 0.5 nm) and therefore suitable as bottom electrodes in heterostructures.

SrRuO$_3$ (SRO) is a conductive and itinerant ferromagnet perovskite with a Curie temperature (T$_C$) of ~160K [1]. It presents a pseudo-cubic lattice with a cell parameter a=0.393nm at room temperature, very close to the cell parameter of SrTiO$_3$ (STO, a=0.391nm). This fact, together with its high chemical stability, makes it ideally suited for application in thin film heterostructures, for example, as a ferromagnetic electrode in magnetic tunnel junctions or as a conductive electrode in ferroelectric or multiferroic devices. A great deal of work has been devoted to the study of the growth mode, microstructure and surface morphology of epitaxial SRO thin films. Most of this work has been carried out on STO substrates with a (001) out-of-plane orientation [2-7]. Different growth mechanisms have been proposed. For instance, Choi et al. found an initial layer-by-layer growth mode followed by a transition to step-flow after the first few monolayers are covered [3], while Sánchez et al. (using annealed and not chemically-treated substrates) observed an initial 3D growth followed by layer-by-layer or step-flow modes, depending on the miscut angle of the substrate [4,5]. Less attention have been paid to the growth of SRO on other substrates such as LaAlO$_3$ [7], as well as to the growth on STO with different orientations other than the (001). Thin films with a (111) orientation present attractive possibilities, for example, in the case of rhombohedral ferroelectrics or multiferroics (i.e. BiFeO$_3$). It is well known that these materials present a spontaneous electrical polarization along the pseudo-cubic (111) direction [8], being therefore the optimal thin film orientation for combining them with other functional oxides in superlattices [9]. SRO is then a very good candidate as bottom electrode in (111)-oriented heterostructures. However, only few reports can be found in the literature dealing with the growth of SRO (111) [10-12]. The reported morphologies display rather rough surfaces [10,11], suggesting that further studies are needed to fully understand and control the growth and microstructure of SRO (111) films.

In the present paper, we investigate the first stages of the growth of SRO on STO (111). Special attention has been paid to the evolution of the surface morphology as a function of the thickness. We will show that the morphology of the films changes from an initial 3D-microstructure to a 2D-one, followed by a further roughening. For thicknesses ranging between 20 and 30nm, a root mean square (RMS) roughness lower than 0.5nm can be obtained.

(111)-oriented SRO thin films were deposited on (111) STO substrates by Pulsed Laser Deposition (PLD with a KrF excimer laser with λ=248nm) assisted by reflective high energy electron diffraction (RHEED). Prior to deposition, substrates with a nominal miscut angle of 0.1° were cleaned in an ultrasonic bath and fired in $O_2$ at 950 °C for 1h. The SRO deposition was performed at 610 ºC in an oxygen pressure of 0.14 mbar. The laser fluence was set to 2.5 $J/cm^2$, with a repetition rate of 1Hz. After deposition, the films were slowly cooled down (-5 ºC/min) to room temperature under an oxygen pressure of 100 mbar. The crystallinity and structure of the films was studied by X-ray diffraction (XRD, Phillips X'Pert MRD diffractometer), while their surface morphology was probed by means of a Nanoscope IIIa atomic force microscope (AFM). The growth rate was estimated as ~0.053Å/pulse. The thickness of the grown films ranged between 1.3 and 46.4nm.

The XRD pattern displayed in Figure 1(a) (Main Panel), corresponding to a 22.4 nm film, shows that the films are epitaxial and (111) oriented. Intensity oscillations due to the finite size of the film in the (111) direction are observed close to the film reflections. The rocking curve of Figure 1(a) (Inset), recorded on the SRO (222) reflection, displays a full width half maximum (FWHM) of 0.07°, reflecting a low degree of mosaicity of the films. The phi-scans of Figure 1(b), performed on the $(110)_{STO}$ and $(400)_{SRO}$ reflections, confirm the cube-on-cube growth of our films. The extracted interplanar SRO(111) distance was $d_{(111)}$=0.228nm for the whole range of explored thickness. This value is slightly larger than the bulk one (0.227nm), suggesting that our films grow under compressive stress. Reciprocal space maps recorded around the (112) reflection, as shown in Figure 1(c) for a 22.4nm film, confirm that the films are fully strained, as the spots corresponding to both the substrate and the film present the same in-plane (parallel) projection of the reciprocal lattice vector. Magnetization measurements as a function of the temperature (Figure 1(d)) showed Curie temperatures very close to the value of the bulk compound (160K [1]).

The AFM image of Figure 2(a) shows the morphology of a treated (1h, 950 °C) STO(111) substrate. The surface displays a flat disordered microstructure, with a typical RMS roughness of 0.17nm. No well defined structure of straight terraces and steps is observed, in contrast to the case of STO(001) crystals under similar annealing conditions [13]. This difference should be related to the polar nature of the (111) surface [14]. Figures 2(b)-(e) show the morphology of SRO(111) films with thicknesses of 1.3, 6.2, 25.5 and 46.4nm, respectively. Figure 2(b) shows that the thinnest film (1.3nm) displays a 3D morphology with the presence of round islands. This is reflected on the transmission spotty RHEED pattern of the corresponding inset. The extracted RMS roughness was 1.3nm. The islands are randomly distributed on the substrate surface, as confirmed by the Fast Fourier Transform (FFT) of the corresponding AFM image, which did not show clear periodicities in any direction. The height profile of Figure 3(a) shows a height variation of ~5nm, while the estimated average lateral size of the islands is ~65nm. It is worth noting that the intensity of the RHEED specular reflection decays immediately when the growth starts, with absence of oscillations, and that the spotty diffraction pattern forms fast, indicating the presence of an initial Volmer-Weber growth [15]. When increasing the thickness of the film, the islands start to coalesce, indicating that the adatom incorporation at this stage of growth happens preferentially in the valleys between islands. The morphology of Figure 2(c), corresponding to a 6.2nm film, shows a structure of meandered flat terraces and valleys with a RMS roughness of 0.95nm. Once the coalescence is complete (Figure 2(d), corresponding to a 25.5nm film), the roughness drops below 0.5nm and the microstructure is composed of flat terraces with an overall height variation lower than 1nm, as shown in the height profile of Figure 3(b). As expected, the corresponding RHEED pattern (inset of Figure 2(d)) switches to a streaky one, typical of 2D surfaces. A new roughening process starts when the thickness is further increased, with the formation of 3D islands, as depicted in Figure 2(e) for a 46.4nm film. The profile displayed in Figure 3(c), corresponding to the same film, covers a height variation of 5nm, and the estimated average lateral size of the islands is ~165nm. Both the height and lateral size of the islands increase with the thickness. The RHEED pattern remains streaky when the thickness is increased, indicating that the 3D structures formed at this stage of growth are wide enough to preclude electron transmission. The progressive roughening with the thickness is reflected in a weakening of the RHEED specular reflection.

The evolution of the RMS roughness in the whole range of explored thickness is collected in Figure 4. After an initial increase in the first stages of growth, the RMS roughness progressively lowers until reaching a value <0.5nm for thicknesses between 20 and 30nm; after that, a further rise starts. The existence of a clear minimum in the RMS roughness is relevant from the point of view of possible applications.

It is worth comparing the evolution of our films morphology against the case of SRO(001) films reported by Sánchez [4-6]. In both cases, the formation of 3D structures is initially observed; however, in the case of (001) films these nanostructures nucleate at the steps of the substrate, leading to self-organized finger-like nanostructures. In our case, as the used STO(111) substrates did not present an ordered structure of steps and terraces, the nucleation takes place at random sites, leading to a disordered island distribution. Another striking difference occurs when the island coalescence is complete and the surface becomes 2D-like. In the case of SRO(001), the growth mode switches concomitantly to either layer-by-layer or step-flow modes, depending on the substrate miscut, the surface remaining flat up to thicknesses higher than 100nm [4,5]. Instead, we observed the formation of 3D islands and a progressive roughening, probably due to an electronic effect related to the polar nature of the (111) surface.

In summary, we have studied the evolution of the morphology of SRO(111) films on STO(111) substrates. We have found that the films morphology changes from an initial microstructure of islands (3D) to one of flat terraces (2D) with increasing thickness; subsequently, for even larger thickness, 3D structures are formed again. For thicknesses between 20 and 30nm, the SRO(111) films present a RMS roughness below 0.5nm, suitable to be used as electrodes in heterostructures.

The authors are grateful to F. Sánchez for the critical reading of the manuscript, to G. Rijnders for useful discussions and C. Daumont for the preparation of the SRO target. This work was supported by the E.U. STREP MaCoMuFi (Contract FP6-NMP3-CT-2006-033221).

Figure Captions

Figure 1: X-ray diffraction characterization performed on a 22.4nm SRO thin film grown on STO(111): (a) (Main Panel) ω-2θ scan; (a) (Inset) Rocking curve (ω-scan), recorded on the $(222)_{SRO}$ reflection; (b) ϕ-scans corresponding to $(110)_{STO}$ and $(400)_{SRO}$ reflections; (c) Reciprocal space map (RSM) recorded around the (112) Bragg peak. $k_0$ is defined as $2\pi/\lambda$, with λ being the X-ray wavelength; (d) Magnetization as a function of the temperature corresponding to a 30nm SRO/STO(111) thin film. The measurement was performed under a field of 1kOe, applied in the plane of the film.

Figure 2: AFM topographic images corresponding to: (a) an STO(111) treated (1h, 950C in $O_2$) substrate; (b)-(e) SRO/STO(111) thin films with thickness of 1.3, 6.2, 25.5 and 46.4nm, respectively. The insets display the RHEED patterns of the corresponding samples.

Figure 3: (a)-(c) Height profiles extracted along the lines displayed on the AFM images of Figures 2(b),(d) and (e), corresponding to 1.3, 25.5 and 46.4nm SRO/STO(111) thin films, respectively.

Figure 4: Root mean square (RMS) roughness as a function of the thickness for SRO/STO(111) thin films with thickness between 1.3 and 46.4nm.

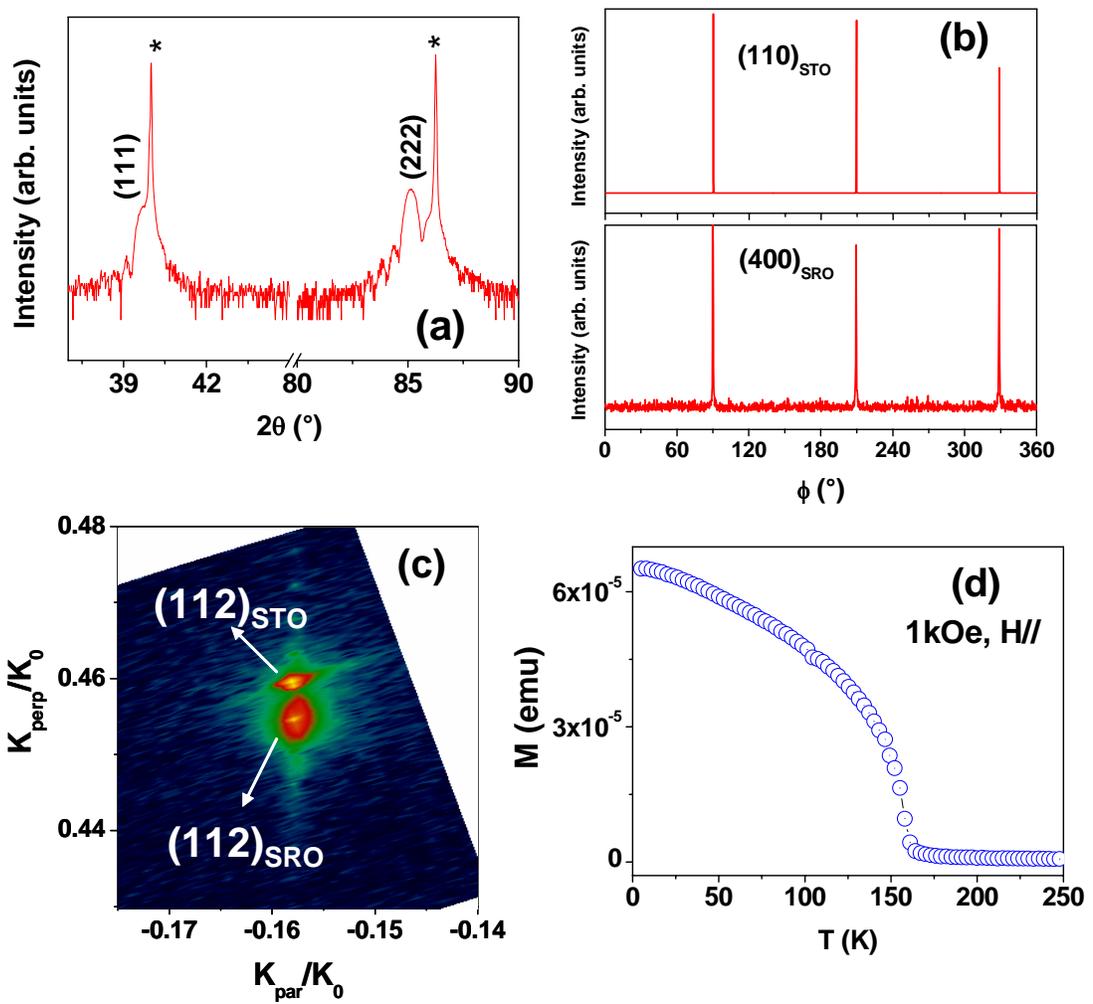

**Figure 1**

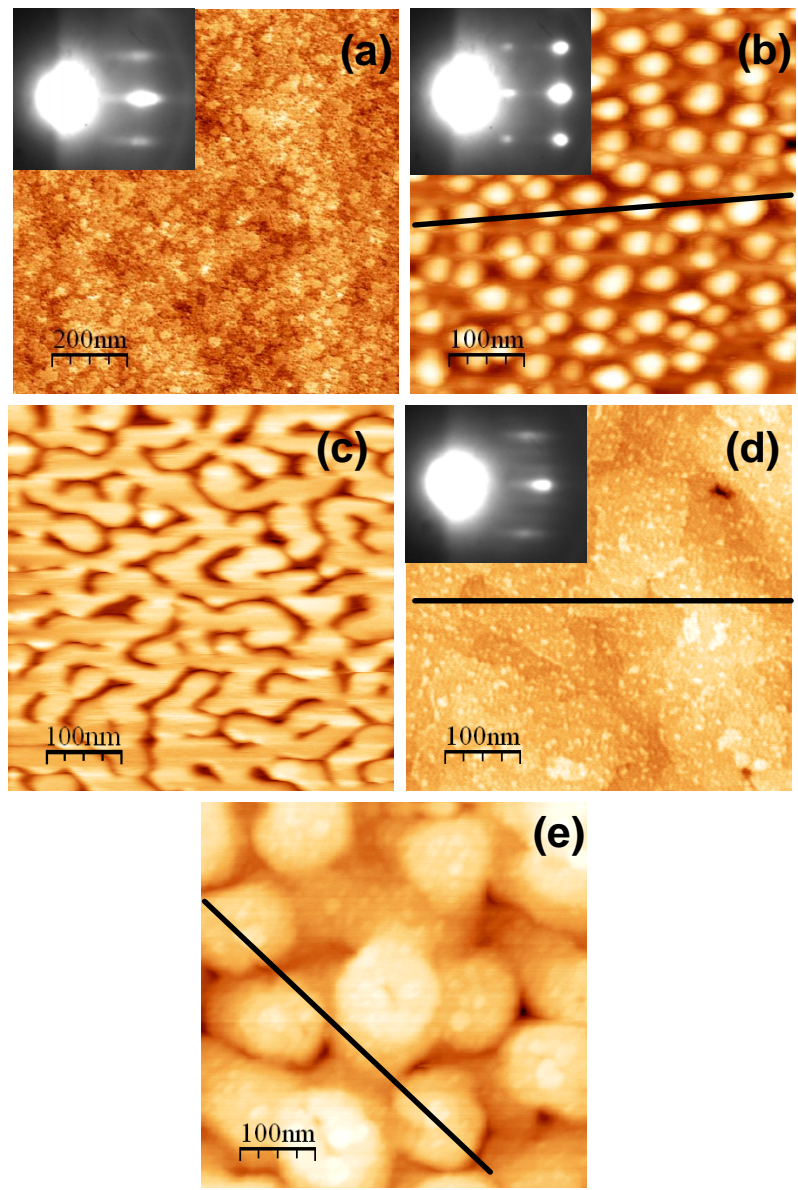

**Figure 2**

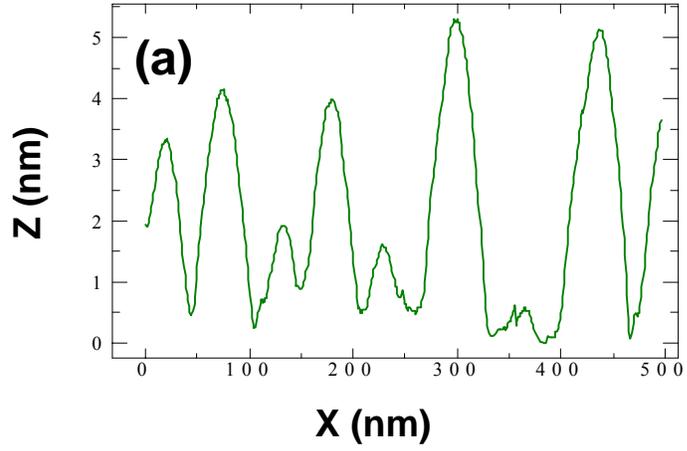
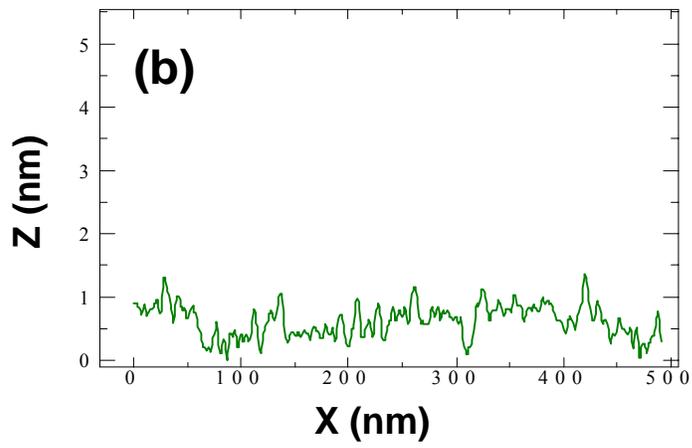
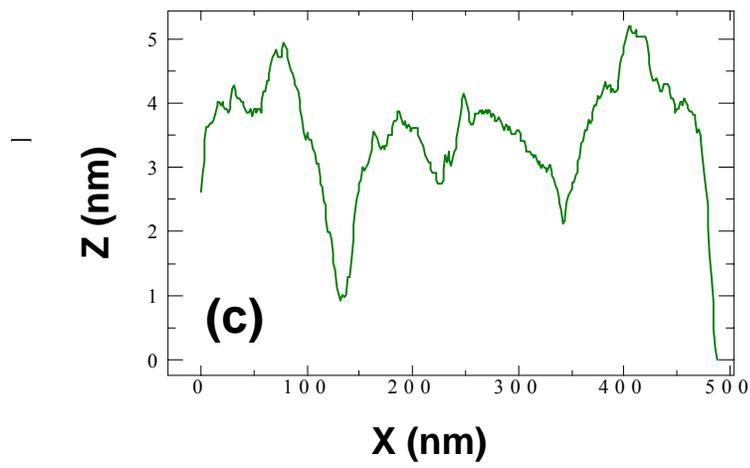

**Figure 3**

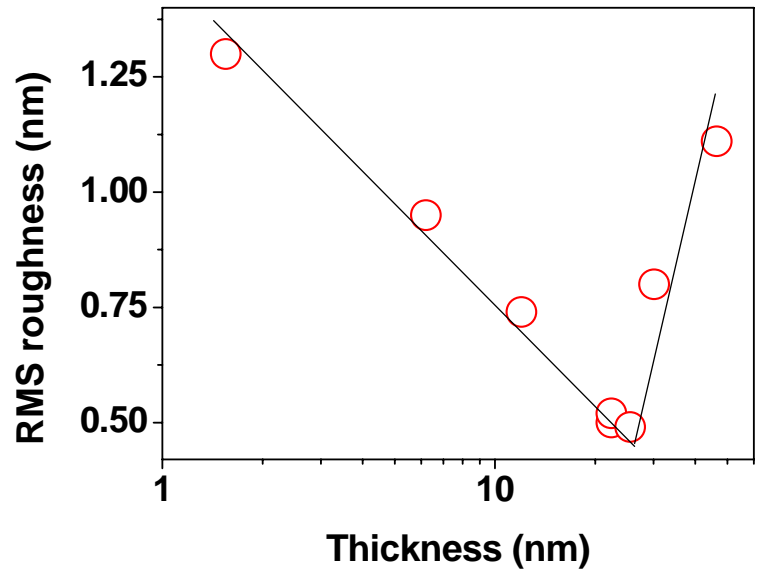

**Figure 4**